\documentclass[12pt]{article}
\usepackage{slashed}
\usepackage{amsmath}
\usepackage{amsfonts}
\usepackage{indentfirst}
\usepackage{color}

\newcommand{\be}{\begin{equation}}
\newcommand{\ee}{\end{equation}}
\newcommand{\bea}{\begin{eqnarray}}
\newcommand{\eea}{\end{eqnarray}}

\newcommand{\nn}{\nonumber\\}

\newcommand{\ol}{\overline}

\begin{document}
 
\begin{flushleft} 
KCL-PH-TH/2013-{\bf 14} \quad LCTS/2013-{\bf 08} 
\end{flushleft}

\vspace{1cm}

\begin{center}

{\Large{\bf Lorentz-Violating Regulator Gauge Fields as the Origin of Dynamical Flavour Oscillations}}

\vspace{1cm}

{\bf Jean Alexandre}$^{(a)}$\footnote{jean.alexandre@kcl.ac.uk}, {\bf Julio Leite}$^{(a)}$\footnote{julio.leite@kcl.ac.uk} 
and {\bf Nick E. Mavromatos}$^{(a,b)}$\footnote{nikolaos.mavromatos@kcl.ac.uk}

\vspace{0.5cm}

\small{
$(a)$ King's College London, Department of Physics, Theoretical Particle Physics and Cosmology Group, London WC2R 2LS, UK \\
$(b)$ CERN, Physics Department, Theory Division, Geneva 23 CH-1211 Switzerland, }

\vspace{1cm}

{\bf Abstract}

\end{center}

\vspace{0.5cm}

We show how a mass mixing matrix can be generated dynamically, for two massless fermion flavours coupled to a Lorentz invariance violating (LIV) gauge field. 
The LIV features play the role of a regulator for the gap equations, and the non-analytic dependence of the dynamical masses, as functions
of the gauge coupling, allows
to consider the limit where the LIV gauge field eventually decouples from the fermions. 
Lorentz invariance is then recovered, to describe the 
oscillation between two free fermion flavours, and we check that the finite dynamical masses are the only effects of the original LIV theory. 
We also discuss briefly a 
connection of our results with the case of Majorana neutrinos in both, the standard model, where only left-handed (active) neutrinos are considered, 
and extensions thereof, with sterile right-handed neutrinos.

\vspace{1cm}

\section{Introduction and Motivation}

The generation of quark, lepton and vector boson masses, as described in the standard model due to their coupling with the 
Higgs boson (Spontaneous Symmetry Breaking), 
seems to have been confirmed by the latest experimental results at the Large Hadron Collider \cite{Higgs}, with the discovery
of a Higgs-like (scalar) particle. 
However, the origin of neutrino masses is still not well established, although the seesaw mechanism seems the most elegant and 
simple for such a purpose~\cite{seesaw}. 
Seesaw mechanisms involve necessarily Majorana mass type fermions, and heavy right-handed neutrino states, without standard 
model interactions (sterile), whose exchange explains the smallness of the active neutrino (left-handed) species of the standard model. 
Such sterile neutrinos have not yet been discovered in Nature~\cite{white2}.

The possibility, therefore, 
of generating neutrino masses dynamically without the involvement of heavy right-handed states is still at play. 
In this article we take some preliminary steps in this direction and envisage scenarios in which flavour oscillations can arise dynamically, 
from the flavour-mixing interaction
of two massless bare fermions with an Abelian gauge field, which has a Lorentz-Invariance-Violating (LIV) propagator.
Lorentz symmetry violation is achieved by higher order space derivatives, which are suppressed by a large mass scale $M$.
This mass scale allows the dynamical generation of fermion masses, as was shown in \cite{JA} with the Schwinger-Dyson approach.
Another role of this mass scale is to lead to a finite gap equation, and therefore to regulate the model. Further studies using a similar model 
were done in \cite{AM1} to generate a fermion mass hierarchy. 

Moreover, LIV U(1) gauge models of the form suggested in \cite{JA} have been shown~\cite{NM} to arise in the low-energy limit of some 
consistent quantum gravity theories, for instance when the U(1) gauge theory is embedded in a stringy space time foam model, with the 
foamy structures being provided by (point-like) D-brane space-time defects (``D-particles''). In such microscopic models, the gauge field 
was one of the physical excitations on brane world universes interacting with the D-particles.
It was observed in \cite{NM} that the LIV Lagrangian of \cite{JA} can be obtained from a Born-Infeld-type Lagrangian of the U(1) gauge field 
in the D-particle background, 
upon an expansion in derivatives. Lorentz Violation arises locally in such models as a result of the recoil of the D-particle defects during 
their interaction with open strings representing the U(1) excitations. 
Other works involving quantization of higher-order-derivative extensions of Quantum Electrodynamics can be found in \cite{mariz}.

An important point is the following structure of the dynamical fermion mass \cite{JA}
\be\label{mdyn}
m_{dyn}\simeq M \, \exp(-a/e^2)~, 
\ee
where $a$ is a positive constant and $e$ is the gauge coupling. Such a non-analytical form is well-known in the studies
of magnetic catalysis \cite{mag}, and it can be derived from a non-perturbative approach only, as the 
Schwinger-Dyson derivation of a gap-equation, used in \cite{JA} and here. 
From the expression (\ref{mdyn}), one can see that it is possible to take the simultaneous limits
\be\label{melim}
M\to\infty \quad {\rm and} \quad e \to 0 ~,
\ee
in such a way that the dynamical mass (\ref{mdyn}) remains finite, corresponding to a \emph{physical} fermion mass.
This procedure is consistent in the string-embedding case of \cite{NM}, where Lorentz symmetry 
is recovered in the limit of vanishing density of D-particles. 
In that model, the LIV scale can diverge in the case of vanishing D-particle density and
zero fluctuations of the recoil velocity (evaluated over a stochastic population of D-particles), where also the coupling can go to zero~\cite{li}, 
in such a way that the
dynamically generated fermion mass remains finite.
This is a physical case in which the vector $U(1)$ fields 
appear as LIV regulators, implying dynamical mass for fermions. 

In the present article we shall consider the regularisation (\ref{melim}) in a more generic sense.
We note at this point that the role of Lorentz symmetry breaking as an UV regulator of a quantum field theory has been considered in \cite{visser}, 
but from a rather different perspective than ours. 
Our aim here is to discuss the dynamically generated mass for fermions and/or the induced oscillations among fermion species, using the coupling of 
the fermions with such a \emph{ LIV regulator gauge field.}

Oscillations of massless neutrinos were already studied in \cite{Benatti}, where neutrinos are considered open systems, 
interacting with an
environment. Such oscillations have also been studied in \cite{masslessLV}, in the framework of LIV models, 
involving non-vanishing vacuum expectation values for vectors and tensors. Other constraints and consequences of these LIV models are given in 
\cite{Altschul}. Whilst these studies have been questioned by phenomenological 
constraints \cite{excluded}, our present model, based on higher order space derivatives, is not excluded.
We note that dynamical generation of flavour oscillations was also studied in the context of Lifshitz theories \cite{Anselmi2}, and 
a detailed analysis of this mechanism was done in \cite{ABH}, for two Lifshitz fermions coupled by a renormalizable four-fermion interaction.

In the limit (\ref{melim}), the non-physical gauge field decouples from the theory, and hence the gauge dependence of the
dynamical mass is avoided (although this problem can be understood perturbatively in 
the framework of the pinch technique~\cite{PT}, as explained in \cite{NM}).
We stress here an essential feature of the mechanism described in the present article. Although LIV operators are suppressed by 
a large mass scale, so that the corresponding effect is negligible at the classical level, quantum corrections completely change this picture, 
and lead to finite effects. In our present study, the finite effect is the dynamical generation of fermion masses, which is present 
even after setting the LIV-suppressing mass scale $M$ to infinity. 
Note that the order of the steps followed is important: quantization is done for finite mass $M$ and coupling $e$, after which the simultaneous 
limits (\ref{melim}) are taken.

The structure of the article is the following:
Next section \ref{sec:2}  introduces the model and derives the corresponding gap equations which must be satisfied by the dynamical masses. We consider the 
corresponding constraints and calculate the dynamical masses in the relevant cases in section \ref{sec:3}. 
In subsection \ref{sec:4} we discuss the ``Lorentz-Invariant limit'' (\ref{melim}),  in which the LIV gauge field 
decouples from fermions, and we demonstrate that relativistic dispersion relations for fermions are indeed recovered.
The extension of the Dirac fermion case to chiral Majorana fermions, as appropriate for neutrinos either in the standard model or in seesaw-type extensions thereof, involving sterile neutrinos, is discussed in section \ref{sec:5}. 
Finally conclusions and outlook are presented in section \ref{sec:conc}. Technical aspects of our work are given in two Appendices.

\section{Dynamical Fermion Mass Matrix \label{sec:2}}

\subsection{The Field Theory Model}

The LIV model we consider is
\be\label{Lag}
\mathcal{L}= -\frac{1}{4} F_{\mu\nu}(1-\frac{\Delta}{M^2})F^{\mu\nu}+\bar{\Psi}(i \slashed{\partial}-\tau \slashed{A})\Psi,
\ee
where $F_{\mu\nu}$ is the Abelian field strength for the gauge field $A^\mu$ and $\Delta=-\partial_i\partial^i$ is the Laplacian 
(the metric used throughout this work is diag(1, -1, -1, -1)). The mass scale $M$ suppresses the LIV derivative operator $\Delta$, 
and can be thought of as the Plank mass, which eventually will be set to infinity. $\Psi$ is a massless fermion doublet
\be
\Psi = \begin{pmatrix} 
       \psi_1 \\
       \psi_2
       \end{pmatrix}~,
\ee
and the flavour mixing matrix $\tau$ features the gauge couplings $(e_1,e_2,\epsilon)$ as 
\be\label{tau}
\tau = \begin{pmatrix}e_1 & -i\epsilon \\i\epsilon & e_2\end{pmatrix}=
\frac{e_1 + e_2}{2}{\bf1} +\frac{e_1 - e_2}{2} \sigma_3 +\epsilon \sigma_2~,
\ee
where $\sigma_i$ are the usual Pauli matrices and ${\bf 1}$ is the $2\times2$ identity matrix. The fermions 
$\psi_1$ and $\psi_2$ in (\ref{Lag}) are Dirac, but the structure of the gap equations that will be derived bellow remains the same in 
the case of Majorana fermions, hence  
the corresponding dynamical masses are independent of the nature of fermions. As already noted in the previous section,  the Lagrangian 
(\ref{Lag}) can be derived from a stringy space time foam model, as shown in \cite{NM}.  We mention in passing that such a space time foam 
model was already used to study decoherence in flavour oscillations, both in flat space time and in a Friedman-Robertson-Walker 
metric \cite{AFMP}.

The gauge field bare propagator is
\be\label{gaugeprop}
D_{\mu\nu} = -\frac{i}{1 +\vec{p}^2/M^2}\left(\frac{\eta_{\mu\nu}}{\omega^2-\vec{p}^2}+\zeta \frac{p_\mu p_\nu}{(\omega^2-\vec{p}^2)^2}\right)~,
\ee
where $\zeta$ is a gauge fixing parameter, which appears in the final expression for the dynamical masses, but do not play 
a role in the simultaneous limits 
\begin{equation}\label{metlim}
M \to \infty \quad {\rm and} \quad e_1, \,  e_2, \, \epsilon \to 0~, 
\end{equation}
that leave the dynamical masses finite, 
as we discuss further on. 
 
We note that the flavour mixing interaction $\ol\Psi\tau\slashed{A}\Psi$ can be at the origin of a gauge boson mass, which is dynamically generated, as 
fermion masses. This alternative to the Higgs mechanism is explained in \cite{dynmassA}, and was extended to a LIV model in \cite{AM2}. In the present 
article, we disregard the possibility to generate a gauge boson mass dynamically, since, as we shall demonstrate below, the flavour mixing coupling $\epsilon$ 
vanishes \emph{necessarily} for consistency of the model in the case there is dynamical generation of \emph{fermion oscillations}. 

The bare fermion propagator is $S=i\slashed{p}/p^2$, where $p_\mu=(\omega,\vec p)$, 
and we assume the dynamical generation of the fermion mass matrix
\be
{\bf M}=\begin{pmatrix} m_1 & \mu \\ \mu & m_2\end{pmatrix}
=\frac{m_1 + m_2}{2}{\bf1} + \frac{m_1 - m_2}{2} \sigma_3 + \mu \sigma_1 ~,
\ee
with eigenvalues
\be\label{eigenmasses}
\lambda_{m\pm} = \frac{m_1+m_2}{2} \pm \frac{\sqrt{(m_1-m_2)^2+4\mu^2}}{2}~.
\ee

Because the mass matrix contains in general non-diagonal elements, the flavour eigenstates $|\psi_i \rangle $, $i=1,2$ are not 
the same as the mass eigenstates 
$|\psi_\pm \rangle$ and there is mixing and oscillations, provided the energy eigenvalues $E_\pm = \sqrt{p^2 + \lambda_\pm^2}$ 
are different~\footnote{We shall check in subsection \ref{sec:4} that the relativistic dispersion relations for the fermions are 
indeed obtained in the Lorentz Invariant Limit (\ref{metlim}), after (finite) dynamical mass generation.}. 
As usual, the flavour eigenstates are connected to the mass (energy) eigenstates by a unitary transformation, parametrised by a mixing angle $\theta$:
\begin{equation}\label{mixed}
\begin{pmatrix} \psi_1 \\ \psi_2 \end{pmatrix} = \begin{pmatrix} {\rm cos}\theta \quad {\rm sin}\theta \\ -{\rm sin} \theta \quad {\rm cos}\theta \end{pmatrix} \, 
\begin{pmatrix} \psi_+ \\ \psi_- \end{pmatrix}
\end{equation}
and if at time $t=0$ one has a flavour $\psi_1(t=0)$ then the probability of obtaining (under Hamiltonian evolution) the other flavour 
$\psi_2(t)$ at $t > 0$ is non trivial and given by 
\begin{equation}\label{osc}
{\mathcal P}_{12} (t) = {\rm sin}^2 \, 2\theta \, {\rm sin}^2 \Big(\frac{E_+ - E_- }{2}\, t \Big)~.
\end{equation}
and the survival probability ${\mathcal P}_{11} = 1 - {\mathcal P}_{12}$. These constitute the \emph{flavour oscillations.} 
We stress that, as becomes evident from (\ref{osc}),  non-trivial mixing, $\theta \ne 0$, is not sufficient for oscillatory behaviour among 
flavours, one needs necessarily different energy levels $E_+ \ne E_- $ as well. 
In what follows we shall identify cases where mixing and/or oscillations are generated dynamically, as a result of the coupling of the fermions 
with the LIV gauge bosons. 

If we neglect other quantum corrections, the dressed fermion propagator $G$, obtained by solving the equation
\be 
G(\slashed p-{\bf M})=i{\bf 1}~,
\ee 
is then
\bea\label{gfermi}
G&=& i\frac{p^2 + \slashed{p}( m_1 + m_2) + m_1 m_2 -\mu^2}{(p^2 - m_1^2)(p^2 - m_2^2)-2 \mu^2(p^2 + m_1 m_2)+\mu^4}\\
&&\times \left[(\slashed{p} - \frac{m_1 + m_2}{2}) {\bf1} + \frac{m_1 - m_2}{2} \sigma_3 +\mu \sigma_1 \right]~.\nonumber
\eea
We must check in what follows that the dynamical masses $m_1,m_2,\mu$ assumed here can indeed be generated by quantum corrections, 
which is obtained using the Schwinger-Dyson approach.

\subsection{Schwinger-Dyson Gap equations}

The self-consistent Schwinger-Dyson equation for the fermion propagator has the usual structure \cite{IZ}, and 
is not modified by the LIV term in the Lagrangian (\ref{Lag}). If we neglect corrections to the wave functions, the vertices and the
gauge propagator, the Schwinger-Dyson equation reads for our model 
\begin{eqnarray}\label{SD}
G^{-1}-S^{-1} = \int_p D_{\mu\nu}~ \tau\gamma^\mu ~G ~\tau\gamma^{\nu}~.
\end{eqnarray}
The previous loop integral is finite as a consequence of the LIV term $\vec p^2/M^2$ in the denominator of the gauge 
propagator (\ref{gaugeprop}). 
We show in Appendix A that the equation (\ref{SD}) 
leads to the following four gap equations, which must be satisfied by the three masses $m_1,m_2,\mu$,
\bea\label{4equs}
\frac{m_1}{4+\zeta} &=& (e_1^2 m_1 + \epsilon^2 m_2) I_1+ (\mu^2-m_1 m_2)(e_1^2 m_2 +\epsilon^2 m_1) I_2\\
\frac{m_2}{4+\zeta} &=& (e_2^2 m_2 + \epsilon^2 m_1) I_1 + (\mu^2-m_1 m_2)(e_2^2 m_1 +\epsilon^2 m_2) I_2\nn
\frac{\mu}{4+\zeta} &=& \mu(e_1 e_2-\epsilon^2)[I_1 - (\mu^2-m_1 m_2)I_2]\nn
0 &=&\epsilon(e_1 m_1 + e_2 m_2) I_1+\epsilon(\mu^2-m_1 m_2) (e_1 m_2 +e_2 m_1) I_2~,\nonumber
\eea
where 
\bea\label{I1I2J}
I_1&=& \frac{J(A_+^2)-J(A_-^2)}{A_+^2-A_-^2}  \\
I_2&=& \frac{1}{A_+^2-A_-^2} \left[\frac{J(A_+^2)}{A_+^2}-\frac{J(A_-^2)}{A_-^2}\right]\nonumber~,
\eea
and 
\bea\label{Apmdef}
J(A_\pm^2)&=&\frac{1}{4 \pi^3}\int_0^{\infty}{dp} \frac{\vec{p}^2}{1 + \vec{p}^2/M^2} \int_{-\infty}^{\infty}{d\omega} 
\left( \frac{1}{\omega^2+\vec{p}^2}-\frac{1}{\omega^2+\vec{p}^2+A_{\pm}^2}\right)\nn
A_\pm^2&=&\frac{m_1^2+m_2^2+2\mu^2}{2} \pm \frac{\sqrt{(m_1^2-m_2^2)^2+4\mu^2(m_1+m_2)^2}}{2}~.
\eea
After the integration over the frequency $\omega$ and the momentum $\vec p$ in the integrals $J(A_\pm^2)$, we obtain for $M>>m_1,m_2,\mu$
\bea\label{I1I2}
I_1&\simeq&\frac{1}{16\pi^2} \frac{1}{A_+^2 - A_-^2} \left[A_-^2\ln\left(\frac{A_{-}^2}{M^2}\right)- A_+^2\ln\left(\frac{A_{+}^2}{M^2}\right)\right]\nn
I_2&\simeq& \frac{1}{16\pi^2}  \frac{1}{A_+^2 - A_-^2} \ln{\left(\frac{A_{-}^2}{A_{+}^2} \right)}~.
\eea 
The four equations (\ref{4equs}) do not have obvious solutions, since they must be satisfied by only three unknowns $m_1,m_2,\mu$.
In what follows, we study different solutions. The ones allowing for the generation of flavour oscillations must have $\mu\ne0$.

\subsection{Constraints}

From the first two equations (\ref{4equs}), one obtains for $e_1^2e_2^2\ne\epsilon^4$: 
\bea\label{I1I2bis}
I_1&=&\frac{1}{4+\zeta}~\frac{e_2^2m_1^2-e_1^2m_2^2}{(e_1^2e_2^2-\epsilon^4)(m_1^2-m_2^2)}\\
(\mu^2-m_1m_2)I_2&=&\frac{1}{4+\zeta}\frac{m_1m_2(e_1^2-e_2^2)+\epsilon^2(m_2^2-m_1^2)}{(e_1^2e_2^2-\epsilon^4)(m_1^2-m_2^2)}~,\nonumber
\eea
and the third and forth equations lead to the following constraints respectively
\bea\label{constraints}
\mu(m_1+m_2)(e_2m_1+e_1m_2)(e_1-e_2)&=&0\\
\epsilon(e_2m_1+e_1m_2)&=&0~.\nonumber
\eea
We are therefore left with different possibilities, that we study in the next section.
Note that, although the denominators in eq.(\ref{I1I2bis}) vanish when $m_1^2=m_2^2$, we will see in the next section that no singularity arises, 
since the numerator then also vanishes, because $e_1=e_2$.

\section{Solutions of the Gap Equations - Dynamical Fermion Masses and Mixing \label{sec:3}}

We now detail the different solutions to the gap equations (\ref{4equs}). The trivial  solution corresponds to the 
situation where no dynamical mass is generated, $m_1=m_2=\mu=0$, which is of no interest to us here.  
In what follows we focus on situations, in which fermion masses are generated with the constraints (\ref{constraints}) satisfied.

\subsection{The case $m_1=m_2=0$ and $\mu\ne0$}

In this case, the eigen masses are
\be
\lambda_\pm=\pm\mu~,
\ee
and the mass eigenstates are 
\be\label{eigenstates}
\psi_\pm=\frac{1}{\sqrt2}(\psi_2\pm\psi_1)~,
\ee
such that the mixing angle (\ref{mixed}) is $\theta =-\pi/4$, in our conventions.\\
This case does not include a mass hierarchy, hence  there are no oscillations (\ref{osc}) among the fermion flavours either, 
since the energy eigenvalues $E=\sqrt{p^2 + \mu^2}$ are the same. 

Among the four gap equations (\ref{4equs}) in this case, only the third is not trivial, and leads to
\bea
\frac{1}{4+\zeta}=(e_1e_2-\epsilon^2)(I_1-\mu^2I_2)~.
\eea
Since $A_\pm^2=\mu^2$, the expressions (\ref{I1I2}) lead to
\bea
I_1&\simeq&\frac{-1}{16\pi^2}\left(1+\ln\left(\frac{\mu^2}{M^2}\right)\right)\\
I_2&\simeq&\frac{-1}{16\pi^2}\frac{1}{\mu^2}~,
\eea
and we obtain
\be
\ln\left(\frac{\mu^2}{M^2}\right)=\frac{-16\pi^2}{(4+\zeta)(e_1e_2-\epsilon^2)}~.
\ee
We note that this expression has a meaning only if $e_1e_2>\epsilon^2$, otherwise $\mu^2>M^2$. 
Assuming this constraint on the couplings, we finally obtain 
\be
\mu\simeq M\exp\left(\frac{-8\pi^2}{(4+\zeta)(e_1e_2-\epsilon^2)}\right)~.
\ee

\subsection{The case $e_2m_1+e_1m_2=0$ and $m_1^2\ne m_2^2$}

In this situation, the first equation (\ref{I1I2bis}) leads to $I_1=0$. The expression (\ref{I1I2}) for $I_1$ leads then to
\be
A_+^2=A_-^2=\exp(-1)M^2~,
\ee
which is not physical, because the dynamical masses are then necessarily of the order $M$. This possibility
is therefore  disregarded, since we will eventually take the limit $M\to\infty$.

\subsection{The case $m_1=-m_2\ne0$ \label{sec:3.4}}

In order to have $m_1=-m_2\equiv m$, it can be seen from eqs.(\ref{4equs}) that
necessarily $e_1=e_2$, such that both constraints (\ref{constraints}) are satisfied. Also, eqs.(\ref{4equs}) are equivalent to
\be
\frac{1}{4+\zeta}=(e^2-\epsilon^2)[I_1-(\mu^2+m^2)I_2]~,
\ee
and $A_\pm^2=m^2+\mu^2$, such that we find
\be
m^2+\mu^2=M^2\exp\left(\frac{-16\pi^2}{(4+\zeta)(e^2-\epsilon^2)}\right)~,
\ee
which has a meaning only if $e^2>\epsilon^2$. This condition allows one to take the limit $\epsilon \to 0$ without 
affecting the mass eigenvalues or mixing angles (see below).
This is important, because, as already mentioned, a non-zero flavour-mixing coupling $\epsilon $ might lead to dynamical 
generation of vector boson masses~\cite{dynmassA,AM2}, thereby spoiling their nature as regulator fields.  

We stress here that we cannot determine $m$ and $\mu$ independently, and the eigen masses are 
\be\label{lpm}
\lambda_\pm=\pm\sqrt{m^2+\mu^2}~.
\ee
The mass eigenstates are
\be
\psi_\pm=\frac{1}{N_\pm}\left(\psi_1+\frac{\mu}{m\pm\sqrt{m^2+\mu^2}}\psi_2\right)~,
\ee
where
\be
N_\pm^2=\frac{2m^2+2\mu^2\pm2m\sqrt{m^2+\mu^2}}{2m^2+\mu^2\pm2m\sqrt{m^2+\mu^2}}~,
\ee
and the mixing angle $\theta$ (\ref{mixed}) is given by 
\be
\tan\theta=\frac{-\mu}{m+\sqrt{m^2+\mu^2}}~.
\ee
In order to fix the mixing angle, one would need an additional ingredient, since the present model does not fix $\mu$, but only $m^2+\mu^2$.

Again, there is no mass hierarchy due to (\ref{lpm}) in this case, the energy eigenvalues are the same, and so 
no oscillations (\ref{osc}) among fermion flavours.

\subsection{The case $m_1=m_2\ne0$: Dynamical Flavour Oscillations \label{sec:3.5}}

We find here from eqs.(\ref{4equs}) that necessarily $e_1=e_2$, $\epsilon=0$ and $\mu^2=m^2$. we have then 
\be\label{I1final}
\mu^2=m_1m_2=m^2~~~\mbox{and}~~~I_1=\frac{1}{(4+\zeta)e^2}~,
\ee
where $e=e_1=e_2$. We have $A_-^2=0$ and $A_+^2=4m^2$, such that the expressions (\ref{I1I2}) 
and (\ref{I1final}) for $I_1$ lead to
\be
-\frac{1}{16\pi^2}\ln\left(\frac{4m^2}{M^2}\right)=\frac{1}{(4+\zeta)e^2}~,
\ee
and the common dynamical mass is finally
\be\label{dynmass}
m=\frac{M}{2}\exp\left(-\frac{8\pi^2}{(4+\zeta)e^2}\right)~
\ee
which, as expected, is not perturbative in $e$. 
In this situation, the mass matrix has identical elements, and has the eigenvalues
\be
\lambda_+=2m=M\exp\left(-\frac{8\pi^2}{(4+\zeta)e^2}\right)~~,~~\lambda_-=0~,
\ee
and the corresponding mass eigenstates are also given by eq.(\ref{eigenstates}). The mixing angle (\ref{mixed}) 
is $\theta = \mp\, \pi/4$, depending on the
sign of $\mu=\pm \, m$, respectively.

In this case, one of the fermions is massless, and the other massive, with mass $2m$. There is a mass hierarchy and thus  
\emph{oscillations} (\ref{osc}) among the fermion flavours in this case. 
We note that because of the constraints (\ref{constraints}), this is the only case in the model (\ref{Lag}) where non trivial 
oscillations among fermion flavours take place. 
As we have seen above, in this case necessarily the flavour-mixing gauge couplings $\epsilon \to 0$, so one does not have to 
worry about dynamical generation of gauge boson masses, and thus the latter play a consistent role as regulator fields.

\subsection{The case $\epsilon=0$, $\mu = 0$}

This is a straightforward generalisation of the original model of \cite{JA}, which involved one fermion, to the two 
fermion-flavour case with no mixing at all.
This case can be divided into two situations: i) $m_1 \ne 0$ and $m_2 \ne 0$, and  ii) $m_1=0$ or $m_2=0$. 
As we shall discuss in section \ref{sec:5}, these may be relevant for Majorana neutrinos in the standard model or extensions thereof, involving right-handed neutrinos, respectively. 

\subsubsection{i) $m_1\ne 0$ and $m_2 \ne 0$ \label{nusm}}

In this first situation, we obtain from (\ref{eigenmasses}) that the two eigenvalues of the mass matrix are
\be\label{2mass}
m_i=M\exp\left(\frac{-8\pi^2}{(4+\zeta)e_i^2}\right)~~~,~i=1,2~,
\ee 
so the dynamically generated mass matrix is diagonal with masses $m_i$ among the two flavours. 
Hence there is no mixing (\ref{mixed}) or oscillations (\ref{osc}) between the flavours $\psi_{i}$, $i=1,2,$ in this case. 

\subsubsection{ ii) $m_1 = 0$ or $m_2=0$\label{sec:3nu}}  

In this case, we observe from the system of equations (\ref{4equs}) that there is also a consistent solution, with either $m_1 =0$ 
with $m_2 \ne 0$ or $m_2 =0$ and $m_1 \ne 0$. The two cases are symmetric. For reasons that will
become clear from our discussion on neutrinos in section \ref{sec:5}, we may concentrate for brevity in the former case, 
i.e. $m_1 =0$. 
In that case, the solution of equations (\ref{4equs}) yields
\be\label{I1b}
I_1 = \frac{1}{(4 + \zeta)\, e_2^2}~,
\ee
while from (\ref{I1I2}) and the definitions (\ref{Apmdef}) we obtain that in this case $A_- =0$, while $A_+^2 = m_2^2 \ne 0$, 
and thus from (\ref{I1b}) we have
\be\label{apm}
|A_+ | = |m_2| \simeq  M ~\exp\left(-\frac{8\, \pi^2 }{(4 + \zeta)\, e_2^2}\right) ~.
\ee
Note that, although $I_2$ diverges logarithmically as $\mu\to0$, it enters the gap equations (\ref{4equs}) only in the combination 
$(\mu^2-m_1m_2) I_2=\mu^2I_2$, which vanishes in this limit.

The mass eigenvalues are in this case, $\lambda_- = 0, $ and $\lambda_+ = m_2 \ne 0$ given by (\ref{apm}). 
The mixing angle $\theta$ is though vanishing and thus there are no oscillations between the states.

\subsection{Lorentz symmetric limit \label{sec:4}}

In order to recover Lorentz invariance, we finally take the simultaneous limits
\be\label{limit}
M\to\infty~~~~\mbox{and}~~~~e_1,~e_2,~\epsilon\to0~,
\ee
in such a way that the dynamical masses are \emph{finite}, and we denote the corresponding ``renormalized'' mass matrix by ${\bf M}_R$.
This procedure is independent of the gauge parameter $\zeta$, and the resulting
fermion mass is set to any desired value. 
In this limit, the gauge field decouples from fermions, and the only finite effect 
from Lorentz violation in the original model is the presence of finite dynamical masses for fermions. 

We now check this statement by demonstrating that the fermion dispersion relations are relativistic in the limit (\ref{limit}). 
We focus here for concreteness on the 
solution described in subsection \ref{sec:3.5}, with $\mu=+m$, but clearly the same conclusion holds for all the other solutions given in section \ref{sec:3}.
Because one of the eigen masses vanishes, which leads to one-loop infrared (IR) divergence, we consider $m_1=m_2=m$ and $m-\mu=m\delta$ with $\delta<<1$.
As will be seen, after the limit (\ref{limit}) is taken, the fermion self energy won't depend on $\delta$, such that the limit $\delta\to0$ will not 
introduce any IR divergence. We calculate in Appendix B the one-loop fermion self energy, where we use the Feynman gauge since the limit (\ref{limit})
is gauge independent.
To lowest order in momentum, we find then
\be
\Sigma=\begin{pmatrix} Z_{diag}^0 & Z_{off}^0 \\ Z_{off}^0  & Z_{diag}^0 \end{pmatrix}\omega\gamma^0
-\begin{pmatrix} Z_{diag}^1 & Z_{off}^1 \\ Z_{off}^1  & Z_{diag}^1 \end{pmatrix}\vec p\cdot\vec\gamma-{\bf M}~,
\ee
where $(\omega,\vec p)$ is the external 4-momentum and
\bea
Z_{diag}^0&=& \frac{e^2}{8\pi^2}\left(\frac{1}{4}-\frac{1}{2} \ln{2} + \frac{1}{2} \ln{\delta} + \ln\left(\frac{m}{M}\right)\right)\\
Z_{diag}^1&=&\frac{e^2}{8\pi^2}\left(-\frac{1}{12}-\frac{1}{2} \ln{2} + \frac{1}{2} \ln{\delta} + \ln\left(\frac{m}{M}\right)\right)\nn
Z_{off}^0&=&Z_{off}^1=\frac{e^2}{16\pi^2}(\ln{2} - \ln{\delta})~.\nonumber
\eea
As expected, because of Lorentz-symmetry violation, $Z_{diag}^0\ne Z_{diag}^1$, but since
\be\label{45}
e^2\ln\left(\frac{m}{M}\right)=-2\pi^2~,
\ee
the limit (\ref{limit}) leads to
\be
\Sigma~~\to~~-\frac{1}{4}(\omega\gamma^0-\vec p\cdot\vec\gamma){\bf 1}-{\bf M}_R~.
\ee
Therefore the dispersion relations are relativistic, since time and space derivatives are dressed with the same corrections in the limit (\ref{limit}).
These corrections can be absorbed in a fermion field redefinition, so that we are left with two free relativistic fermion flavours 
oscillating\footnote{The identity (\ref{45}) is valid in the Feynman gauge, and corrections to the fermion kinetic term are actually 
gauge-dependent. But since these are finite, a redefiniton of coordinates will leave the final Lagrangian gauge-independent.}.

\subsection{Energetics Arguments}

Among the different possibilities to generate masses dynamically, one can question the preference for the system to have non-vanishing masses,
rather than no dynamical mass generated. We give here an energetics argument supporting the choice of non-vanishing dynamical masses~\cite{AM2}.
This argument is based on 
the Feynman-Hellmann theorem~\cite{feynman}, which states that, if there is a ground state $|\Psi_\lambda\rangle $ of a 
system with Hamiltonian that depends on a parameter $\lambda$, then for the energy $E$ of this ground state we have:
\begin{equation}\label{fhth}
\frac{\partial E}{\partial \lambda} = \langle \Psi_\lambda | \frac{\partial \widehat{H}}{\partial \lambda} | \Psi_\lambda \rangle~,
\end{equation} 
where $\widehat{H}$ is the Hamiltonian operator of the system. In our situation, the parameter $\lambda$ can be chosen to be $\lambda=M^{-2}$,
such that
\begin{equation}\label{fhthours}
\frac{\partial E}{\partial \lambda } = +\frac{1}{4}~ {_M }\langle 0 | \int d^4 x^E \left(F_{\mu\nu} \Delta F^{\mu\nu} 
\right)_E | 0 \rangle_M ~, \quad \lambda = M^{-2} ~.
\end{equation}
where the index $E$ denotes Euclidean formalism, as a result of the fact that the Hamiltonian  of the system is identified with minus the 
effective Euclidean action. 
One should expect that the Lorentz-violating nature of the vacuum $|0\rangle_M$ implies in general the non vanishing of the right-hand-side, 
implying a dependence of the vacuum energy on the dynamically generated mass.
Using the cyclic Bianchi identity for the gauge bosons field strengths, 
\begin{equation}\label{bianchi}
\partial_{\left[\mu\right.} F_{\left. \nu\rho\right]} = 0~,
\end{equation}
with the symbol $[ \dots ]$ denoting symmetrisation of the appropriate indices, we obtain
\be
\frac{\partial E}{\partial \lambda } =- \frac{1}{4}~ {_M }\langle 0 | \int d^4 x^E \left(
F_{\mu\nu}\partial_i[\partial^\mu F^{\nu i}+\partial^\nu F^{i\mu}]\right)_E | 0 \rangle_M ~.
\ee
Integrating by part and assuming that the fields decay away at space-time infinity, one may write eq.(\ref{fhthours}) in the form:
\be\label{fhthours2}
\frac{\partial E}{\partial \lambda } = +\frac{1}{2} ~{_M} \langle 0 | \int d^4 x^E \left(\partial^\mu F_{\mu\nu} \partial_i F^{\nu i} 
 \right)_E | 0 \rangle_M
\ee
We write then the equations of motion for the vector fields, from the Lagrangian (\ref{Lag}) where we neglect the operator $\Delta/M^2$,
and we obtain:
\be\label{squarecharge}
\frac{\partial E}{\partial \lambda } 
= +\frac{1}{2} ~{ _M} \langle 0 | \int d^4 x^E \left((J^0)^2+\vec J\cdot\vec J -J_k\partial_0F^{k0}\right)_E|0\rangle_M~,
\ee
where the current is $J^{\mu} = \overline{\Psi} \gamma^\mu \tau \Psi$ 
and the Euclidean formalism is used in (\ref{squarecharge}).
In the framework of the LIV model studied here, one might face a situation where non-trivial condensates of  the 
covariant square of the stationary four-current $J^\mu$ are observed in the (rotationally invariant) vacuum. For such stationary 
currents, where $\partial_0 F^{k0}=0$, we have then from eq.(\ref{squarecharge}):
\begin{eqnarray}\label{monot}
\frac{\partial E}{\partial \lambda } = \frac{1}{2}~_M\langle 0 | \int d^4 x^E \left( J^\mu J_\mu \right)_E |0\rangle_M \ge 0~.
\end{eqnarray} 
This implies that the vacuum energy $E$ in this case 
is a monotonically decreasing function of $M^2$, so that the energy goes to its minimum in the Lorentz symmetric limit $M\to\infty$ 
we are interested in.

The above argument in favour of the stability of the Lorentz Invariant Limit (\ref{melim}) can be turned into an argument  in favour also of the dynamical fermion-mass generation as follows: 
in a finite $M < \infty$ situation, the gauge coupling $e$ is considered as an independent quantity from $M$, and thus, in view of (\ref{mdyn}), 
the LIV mass scale is proportional to the fermion mass $m > 0$ (absolute value if $m < 0$). In this sense, one obtains from (\ref{monot}),
\be\label{mono2} 
\frac{\partial \, E}{\partial \, m} = \frac{\partial \, M}{\partial m} \, \frac{\partial \lambda}{\partial M}\, \frac{\partial E}{\partial \lambda} 
= - \frac{1}{m\, M^2} \, _M\langle 0 | \int d^4 x^E \left( J^\mu J_\mu \right)_E |0\rangle_M \le 0~.
\ee
Thus, the energy of the vacuum for any finite value of $M$ is also a monotonically decreasing function of the fermion mass. 
In the Lorentz-symmetric limit (\ref{melim}), 
 the energy exhibits a plateaux, as far as its dependence on the finite $m > 0$ is concerned, \emph{i.e.} $\partial E/\partial m =0$, 
but its value is lower than the case 
where $m=0$.  

We stress, however, that the above arguments rely on the formation of condensates for the covariant square of the current. 
The latter property is at present a conjecture, and its proof goes far beyond our considerations in this article.

\section{Extension to Chiral Majorana Neutrinos \label{sec:5}}

Above we considered Dirac non-chiral fermions. However, if we wish to present the above-described dynamical mass generation 
scenario as a viable alternatives to standard seesaw mechanisms for neutrinos, and explain the neutrino oscillations as a dynamical phenomenon,  
then we should extend the above considerations to the case where the fermions are chiral and Majorana (as most likely is the case realised in nature). 

Below we shall consider two separate cases. The first is the one in which the fermions correspond to Majorana mass eigenstates obtained from  the left-handed flavour neutrino physical fields of the standard model, while the second case involves sterile right-handed neutrinos as in seesaw extensions of the standard model. 

We shall discuss a connection of our previous findings on dynamical mass generation to both types of neutrino masses.
In particular, we shall first review the underlying formalism, which is necessary for a better and more complete understanding of 
the details of such connections. More specificially, as explained bellow, it is because Majorana fermions are {\it mass} 
eigenstates, involving \emph{both} chiralities,  that our results can be relevant to neutrino oscillations. 
In what follows, we shall first link our dynamical mass generation scenario described in section \ref{nusm} to 
the standard model left-handed neutrinos, and then we shall connect the dynamical mass generation scenario in section \ref{sec:3nu} 
to a dynamical see-saw model, involving right-handed Majorana neutrinos that exist in extensions of the standard model. 
Since in our scenarios the values of the mass can be fixed phenomenologically, we can assume that any 
other mass contributions to neutrinos (\emph{e.g.} due to a Higgs mechanism in conventional see-saw models) are subdominant. The advantage 
of our dynamical mass generation approach lies specifically to the possibility of being applied directly to left-handed standard model 
neutrinos, without the need of introducing right-handed ones (although there may be other reasons to introduce the latter, and this is why in this section we describe both cases).

\subsection{Left-handed Neutrino Majorana Mass Generation}

For instructive purposes it is useful first to review some basic formalism. According to the standard theory~\cite{bilenky} a (Majorana (M)) 
mass term for neutrinos, which involves only left-handed fields, reads
\be\label{majmass}
{\mathcal L}^M = -\frac{1}{2} \overline{\nu_L} M^M \, (\nu_L)^c + {\rm h.c.}~, 
\ee
where the normalisation of 1/2 will be understood in what follows. 
In the one generation case we focus upon here $M^M$ is a c-number (In case of many generations, $\nu_\ell$, $\ell=e,\mu, \tau$ then $M^M$ is a symmetric 
$3 \times 3 $ matrix, as can be seen easily).
The (physical) Majorana field $\nu^M$, involving \emph{both} chiralities,  is defined as 
\be\label{majfield}
\nu^M = \nu_L + (\nu_L)^c 
\ee
and is always an \emph{eigenstate of the mass}, that is when expressed in terms of it the Mass matrix is diagonal
\be
{\mathcal L}^M =  -\frac{1}{2} \overline{\nu^M} M^M \, \nu^M  = - \frac{1}{2}  \sum_{i=1}^3 m_i \, {\overline \nu}_i \, \nu_i ~,
\ee
with $m_i$ the mass eigenvalues. It satisfies the  \emph{Majorana condition}
\begin{equation}\label{majcond}
(\nu^M)^c = \nu^M~,
\end{equation}
which implies that a Majorana field is its own antiparticle. 

The kinetic (Dirac) term of the Lagrangian with respect 
the left-handed  $\nu_L$ fields, 
when expressed in terms of the Majorana mass eigenstate fields $\nu_i$ reads (up to an irrelevant total derivative):
\bea\label{lkin}
{\mathcal L}_{\rm kin} = {\overline \nu}_L i \slashed{\partial} \nu_L &=& 
\sum_j \frac{1}{2} \overline{\nu}_j i \slashed{\partial} \nu_j\\ 
\mbox{with}~~\nu_j &=& (\nu_j)^c = \nu_{j\,L} + (\nu_{j\,L})^c~,\nonumber
\eea
with the suffix $j$ denoting mass eigenstate fields. In view of the extra $\frac{1}{2}$ normalisation of the kinetic terms of the Majorana 
fields then, it is customary to define the corresponding mass terms with the same normalisation, as we have done above, compared to other 
Dirac fields one encounters in the standard model. 

In our approach we consider the coupling of a doublet of (mass eigenstate) Majorana fields to the regulator U(1) gauge field $A_\mu$ in the case discussed in subsection 
\ref{nusm}. The fact that a Majorana field contains both chiralities allows for a straightforward extension of the Dirac case discussed in previous sections to the current situation.
In this way, we are able to generate dynamically different mass eigenvalues for the two species, \emph{without mixing}, 
as implied by the corresponding solutions 
\be
m_i=M\exp\left(\frac{-8\pi^2}{(4+\zeta)e_i^2}\right)~~~,~i=1,2~.
\ee 
This is a consistent way of discussing the dynamical 
appearance of a Majorana mass for left-handed neutrinos of the standard model. Non-trivial mixing of flavour neutrinos, 
coupled to the physical $SU(2)_L$ gauge fields of the standard model, 
can then be obtained in the case where the mass eigenvalues are different. In order to recover the Lorentz symmetric limit in this case, 
we need to take simultaneously 
$e_1, e_2 \to 0$ in such a way that their ratio is fixed to the phenomenologically desired value. 
It is important that in this approach we started from Majorana mass eigenstates coupled to the regulator gauge fields, with no mixing. 
The latter is obtained 
when one expresses the Majorana mass eigenstates in terms of the \emph{flavour} neutrino eigenstates, which appear in nature.

\subsection{Extensions of the standard model with right-handed (sterile) neutrinos}

When there are right-handed (sterile) neutrino components present, $\nu_R$, one can define two kinds of mass terms, Majorana 
(\ref{majmass}) (M) and Dirac (D). 
The most general mass term, then, reads~\cite{bilenky}:
\be\label{MD}
{\mathcal L}^{M+D}= -\frac{1}{2} {\overline \nu}_L \, M_L^M \, (\nu_L)^c - \overline{\nu}_L \, M^D \, \nu_R  -  
\frac{1}{2} {\overline \nu}_R \, M_R^M \, (\nu_R)^c + {\rm h.c.}~,
\ee
where the mass matrices $M^M_{L,R}$ and $M^D$ are in general different.  

We consider below the mixed mass terms (\ref{MD}) in the one generation case, of relevance to our models discussed in this work. 
In this case we may assembly the left-handed  neutrino fields and the conjugate of the right-handed one into a left-handed doublet field
\be\label{nfield}
n_L = \begin{pmatrix} \nu_L \\ (\nu_R)^c \end{pmatrix}
\ee
in which case the mass term (\ref{MD}) can be written in terms of a $2 \times 2 $ mass matrix ($6 \times 6$ in the case of three generations):
\be\label{nMD}
{\mathcal L}^{MD} = -\frac{1}{2} {\overline n}_L \, M^{M+D} \, (n_L)^c ~, \qquad M^{M+D} = \begin{pmatrix} M_L^M \quad M_D \\ (M_D)^T \quad M_R^M \end{pmatrix}
\ee
where for the sake of generality we expressed here the mass matrix as a matrix with flavour components as well. For a single generation of neutrinos, 
we consider below, the elements of the above (2 $\times $ 2 in this case) matrix are c-numbers. 
For our toy purposes here we assume no CP violation in the lepton sector~\cite{bilenky}.

The matrix $M^{M+D}$ can be diagonalised by a Hermitean matrix $U$:
$$ M^{M+D} =  U \, \tilde{\mathfrak m} \, U^T = O \tilde{\mathfrak m} \eta \, O^T $$
with~\cite{bilenky}:
\be\label{Umix}
U = O\, {\eta}^{1/2}~, \qquad O = \begin{pmatrix} {\rm cos}\, \theta \quad {\rm sin} \, \theta \\ -{\rm sin}\, \theta \quad {\rm cos}\, \theta \end{pmatrix} ~,
\ee
where the matrix $\eta $ has eigenvalues $\eta_i = \pm 1 $, which are related to the so-called CP parity of the Majorana neutrinos~\cite{bilenky}, 
and stem from the fact that the mass eigenvalues can be positive or negative 
\be\label{frakmp}
{\mathfrak m}^\prime_{1,2} = \frac{1}{2} (M_R + M_L) \mp \frac{1}{2} \, \sqrt{ (M_R - M_L)^2 + 4 M_D^2 }
\ee
so one can rewrite them as ${\mathfrak m}_i = |{\mathfrak m}_i | \eta_i \equiv \tilde{\mathfrak m}_i \, \eta_i, \, \eta_i = \pm 1, \, i=1,2.$ 
The mixing angle $\theta$ being such that:
\be\label{mangle}
{\rm cos}2 \theta = \frac{M_R - M_L}{\sqrt{(M_R - M_L)^2 + 4 M_D^2 }}~, \quad 
{\rm tan}2\theta = \frac{2 M_D}{M_R - M_L}~.
\ee
The Majorana fields, involving \emph{both} chiralities, are then defined in terms of $U$ as
\be\label{finalmaj}
\nu^M = U^\dagger \, n_L + (U^\dagger \, n_L)^c = \begin{pmatrix} \nu_1 \\ \nu_2 \end{pmatrix} ~, \quad \nu_i^c = \nu_i, \, \quad i=1,2~.
\ee
These are the mass eigenstate fields with masses ${\tilde {\mathfrak m}}_{1,2}$ (\ref{frakmp}).

The original chiral (left-handed) neutrinos, appearing in the Lagrangian (\ref{majmass}) are related therefore to these mass eigenstates as follows:
\begin{eqnarray}\label{etanu}
\nu_L &=& {\rm cos}\, \theta \sqrt{\eta_1} \nu_{1\, L} + {\rm sin}\, \theta \sqrt{\eta_2} \, \nu_{2 \, L}  \nonumber \\
(\nu_R)^c &=& - {\rm sin}\, \theta \, \sqrt{\eta_1} \, \nu_{1\, L} + {\rm cos} \, \theta \sqrt{\eta_2} \, \nu_{2\, L}
\end{eqnarray} 
In the standard seesaw scenarios~\cite{seesaw}, there are no masses for the left handed fields, $M_L = 0$, and the right-handed neutrino (sterile) 
Majorana masses are assumed to be much heavier than the Dirac masses , $M_R \gg M_D$, the latter being given by means of a Higgs mechanism by, \emph{e.g}., 
Yukawa coupling terms of the form
\be\label{yuk}
F\, \phi^C \overline{\nu}_L \, \nu_R + {\rm h.c.},  
\ee
where $F$ is the Yukawa coupling and $\phi^c = i \sigma_2 \phi^\star $ is the dual of the Higgs doublet. In this limit, from (\ref{frakmp}), 
(\ref{mangle}) the mass eigenstates generated are of the form
${\mathfrak m}_1 \simeq \frac{M_D^2}{M_R} \ll M_D$ and ${\mathfrak m}_2 \simeq M_R \gg M_D$, while the mixing angle $\theta \simeq \frac{M_D}{M_R} \ll 1$, 
and also 
$\eta_1 = -1$, $\eta_2 = 1$, hence from 
(\ref{etanu}) we do obtain: 
\bea\label{seesaw} 
  \nu_L &\simeq&  i \, \nu_{1\, L} + \frac{M_D}{M_R} \, \nu_{2\, L} \nonumber \\
(\nu_{R})^c & \simeq&  - i \frac{M_D}{M_R} \nu_{1\, L} + \nu_{2\, L}~. 
\eea

The purpose of the remainder of this section is to adopt the previous procedure and generate \emph{dynamically} masses for the Majorana fields by 
coupling them to gauge fields. 
To this end we view one of the flavours as a right-handed sterile neutrino, $N_R = \frac{1}{2} \Big(1 + \gamma_5 \Big)\, N$ and the other flavour 
$\psi_L = \frac{1}{2} \Big(1 - \gamma_5 \Big)\, \psi$, as an active neutrino of the standard model. Here, 
$N, \psi$ are non-chiral spinors, which in the case of neutrino may be taken to be Majorana. 
 This would be a toy (with one active and one sterile neutrino) version  of the minimal (non supersymmetric) extension of the standard model of 
 ref. \cite{numsm}, termed $\nu$MSM. In this case we do not avoid right-handed neutrinos but we use the dynamical mass generation mechanism 
 presented here to give masses to them. 
In this case the Lagrangian (\ref{Lag}) is replaced by:
\bea\label{Lag3}
\mathcal{L}&=& -\frac{1}{4} F_{\mu\nu}(1-\frac{\Delta}{M^2})F^{\mu\nu}+ 
\bar{N}(i \slashed{\partial}-e_1 \slashed{A})\, \frac{1}{2} \Big(1 + \gamma_5 \Big)\, N \\
&&~~~~~~~~~~~~~~~~~+\bar{\psi}(i \slashed{\partial}- e_2 \slashed{A})\, \frac{1}{2} \Big(1 - \gamma_5 \Big)\, \psi~,\nonumber
\eea
Notice that in this case, due to the opposite chiralities of the two spinor fields, the off diagonal flavour mixing gauge couplings $\epsilon $ 
are irrelevant because the corresponding 
terms vanish identically.

According to our general discussion on combined Dirac and Majorana masses above, we may express this Lagrangian in terms of Majorana fields 
and view the initially massless $\psi$ and $N$ as the Majorana 
field doublet  $\nu^M$ (\ref{finalmaj}), $\nu^M = \begin{pmatrix} \psi \\ N \end{pmatrix} $,
 which then couples to the vector fields. Dynamically generated mixing of the two should involve a small 
mixing angle in phenomenologically realistic situations in view of the discussion above,  \emph{cf}.  Eq.~(\ref{seesaw}).

Unfortunately, in our single gauge field toy models considered here, the only solution from the cases discussed in section \ref{sec:3} that can 
be carried over to the case of Majorana neutrinos is the one discussed in subsection \ref{sec:3nu}. In this case, the masses $m_1,m_2$
can be identified with the dynamically generated mass eigenvalues
\bea\label{dgm}
m_1&=&\lambda_-=0\\
m_2&=&\lambda_+=M\exp\left(\frac{-8\pi^2}{(4+\zeta)e_2^2}\right)\nonumber~,
\eea
where $m_1$ can be identified with the left-handed Majorana mass $M_L=0$, which in the usual seesaw models is assumed zero,
and $m_2$ is then identified with the heavy right-handed Majorana mass, $M_R$. In this way, the dynamically generated masses (\ref{dgm}) 
correspond to a see-saw type mass matrix (\ref{nMD}) of the form ($2\times 2$ in our one-generation example considered explicitly here):
\be\label{nMD2}
M^{M+D} = \begin{pmatrix} 0 \quad 0 \\ 0 \quad m_2 \end{pmatrix}
\ee
for the Majorana neutrinos. 
The reader should note that there is no non-trivial Dirac mass $\mu$, since the latter vanishes in the dynamical solution, as explained in 
subsection \ref{sec:3nu}. 

Nevertheless, the latter can be generated through the usual Yukawa couplings (\ref{yuk})  with the Higgs field, which upon acquiring a vacuum expectation value 
via the Higgs mechanism would generate a Dirac mass term, as we shall discuss below.  In this scenario it is the (heavy) right-handed 
mass that can be generated dynamically, due to the coupling with the LIV gauge sector. Since, as we have already mentioned, the finite 
mass in the Lorentz Invariant limit (\ref{metlim}) is arbitrary, we can arrange so that the latter is much heavier than the Higgs-generated 
Dirac mass, which leads to naturally light active neutrinos in the standard model sector. Let us now proceed to discuss in some detail this latter scenario.

In this case we can consider the Schwinger-Dyson equations in the background of a Higgs field~\footnote{Any contributions of the fluctuations
of the Higgs to the Schwinger-Dyson equations will be suppressed  by the Higgs mass and will be ignored to our leading approximation adopted here.} 
acquiring a v.e.v. $\langle \phi \rangle = v$. This will yield a bare Dirac mass term of the form $Fv$, where $F$ is the pertinent Yukawa coupling. 
This affects the form of the bare fermion propagator $S$ by Dirac-mass terms proportional to 
the Higgs-induced $\mu_0 = F v$, while the dressed fermion propagator $G$ will have a form similar to that in (\ref{gfermi}), but with 
the replacement of $\mu$ by the sum  $\mu + \mu_0$, with $\mu$ the dynamically generated Dirac mass term. It can be readily seen then that the pertinent 
Schwinger-Dyson equations read:
\bea\label{sdmn}
I_1&=&\frac{1}{4+\zeta}~\frac{e_2^2m_1^2-e_1^2m_2^2}{(e_1^2e_2^2-\epsilon^4)(m_1^2-m_2^2)}\\
((\mu + \mu_0)^2-m_1m_2)I_2&=&\frac{1}{4+\zeta}\frac{m_1m_2(e_1^2-e_2^2)+\epsilon^2(m_2^2-m_1^2)}{(e_1^2e_2^2-\epsilon^4)(m_1^2-m_2^2)}~,\nonumber
\eea
supplemented by  the following constraints, similar to those given by eqs.(\ref{constraints}):
{\small \bea\label{constraints2}
(m_1+m_2)\, \Big[ \mu \, (e_2m_1+e_1m_2)(e_1-e_2) & - & \mu_0 \, 
\Big( m_1 \, (\epsilon^2 + e_2^2 ) - m_2 \, (\epsilon^2 + e_1^2) \Big) \Big] = 0 \nonumber \\
\epsilon(e_2m_1+e_1m_2)&=&0~,
\eea}
where we stress once again that $\mu$ is the dynamically generated Dirac mass term, and $\mu_0 = F v$ is the bare (Higgs-induced) one. 
The integrals $I_i$, $i=1,2$ are given by 
the same expressions as in (\ref{I1I2}) but with the replacement of $\mu$ by $\mu+ \mu_0$. 

For consistency with our considerations above, we seek solutions of (\ref{sdmn}) in which $m_2 \ne 0$ and $m_1 = \mu = \epsilon  = 0$, 
which on account of the constraints (\ref{constraints2}) imply $e_1 = 0$. We also make the physically relevant assumption that the Dirac 
mass $\mu_0 \ll m_2 $ (which is consistent with light active neutrino species). To leading order in $x \equiv \frac{\mu_0}{m_2} \ll 1$, we then obtain 
$$ I_1 \simeq \frac{-1}{16\, \pi^2}\ln\left(\frac{m_2^2}{M^2}\right) +{\mathcal O}(x^2) ~~,~~~~ 
\mu_0^2 \, I_2 \simeq {\mathcal O}(x^2\ln x)~.$$

The solution of eqs.(\ref{sdmn}), then, for the dynamically generated mass matrix of the Majorana neutrinos is  
the same as in (\ref{dgm}) but with  the mass matrix having bare $\mu_0 = F v$ Dirac terms, 
\be\label{nMD3}
M^{M+D} = \begin{pmatrix} 0 \quad Fv \\ Fv \quad m_2 \end{pmatrix}~, \, \,  F v \ll m_2~,
\ee
with $m_2$ given by (\ref{dgm}). 
So our dynamical mass generation scenario provides a novel way for generating heavy right-handed neutrino masses when applied to extensions of 
the standard model containing such states, such as the model of Ref.~\cite{numsm}.

\section{Conclusions and Outlook \label{sec:conc}}

In this work we have considered  the coupling of flavoured fermion fields to LIV vector gauge bosons, with Lorentz Invariance being violated in 
the gauge sector at a mass scale $M$ and studied the limiting case where the gauge couplings go to zero, while the LIV Mass scale $M \to \infty $ 
simultaneously, in such a way that the Schwinger-Dyson dynamically generated fermion masses remain finite. No vector boson mass is generated due 
to an appropriate arrangement of the couplings. In this way, the LIV vector bosons are viewed as regulator fields, with the only remnant of the LIV 
the dynamical fermion mass. Unfortunately, the dynamical equations are sufficiently restricted so as to allow 
only one case where oscillation among fermion flavours is allowed and in this case one of the fermion mass eigenstates is massless, while the other is massive. 
The mixing angle is necessarily maximal in this case $\theta = \pm \pi/4$. 
One may hope that extension to a third flavour may lead to more phenomenologically realistic situations with arbitrary mixing and mass generated for all flavours.

Another possibility towards this result might be the inclusion of more than one regulator vector fields, along the lines of \cite{AM2}, where, 
however, not only a mass hierarchy is generated between the fermions, with non trivial masses, but also one of the gauge bosons acquires a mass. 
In our case of regulators, unfortunately, this last mass would also be kept finite, but probably this would not be a problem, since the massive 
vector field  decouples from the Lagrangian of the fermions in the zero gauge coupling ``relativistic limit'' (\ref{metlim}). We hope to come back 
to this case in a future work.

Another aspect of our work, which was also the original motivation, is the one in which this method applies to chiral neutrinos of the standard model, 
in an attempt to discuss neutrino mass generation independently of the seesaw mechanism. We have discussed two scenarios in this respect.

In the first, we avoided the inclusion of sterile neutrinos altogether. In this case the two flavours considered above have been viewed as 
corresponding to Majorana mass eigenstates of two left-handed  
neutrino flavours, interacting with a LIV regulator gauge field with vanishing couplings.  It was demonstrated that different mass eigenstates could 
be obtained in the Lorentz symmetric limit, which then leads to standard oscillations among the physical neutrino flavours coupled to the $SU(2)_L$ gauge fields of the toy standard model involving only two flavours. Extension to the physical case of three generations, including CP violation in the lepton sector,  will constitute the subject of a forthcoming publication. 

The second scenario,  involved 
an extension of the model (\ref{Lag}) to a toy version of the $\nu$MSM model of \cite{numsm}, in which one of fermion flavours of (\ref{Lag}), say 
$\psi_1$, represented a right-handed neutrino field, and the other flavour $\psi_2$ a left-handed active neutrino of the standard model.
In such a case our aim was to generate dynamically a mass hierarchy between active and right-handed (possibly sterile) neutrinos of the type needed 
in phenomenological approaches to dark matter, where a keV sterile neutrino may play the role of a dark matter field, consistently with current 
astrophysical and cosmological data~\cite{numsm}. 
In the context of our framework, we can only generate dynamically the right-handed neutrino mass, but not a Dirac mass term. It is interesting that 
the absence of a left-handed Majorana mass (standard assumption in seesaw models) appears naturally in our models. A Dirac mass then, coupling left(active) 
and right-handed(sterile) components can be generated by the standard Higgs mechanism of the standard model.

\vspace{1cm}

\section*{Acknowledgements} 

The work of J. L. is supported by the National Council for Scientific and Technological Development (CNPq - Brazil), while that of N.E.M. is supported in part  by the London Centre for Terauniverse Studies (LCTS), using funding from the European Research
Council via the Advanced Investigator Grant 267352, and 
by STFC UK under the research grant ST/J002798/1.

\appendix
\section{Appendix: gap equations}

The aim of this appendix is to present the main steps to obtain (\ref{4equs}) from the Schwinger-Dyson equation (\ref{SD}) which is rewritten below:
\begin{eqnarray}\label{SDap}
G^{-1}-S^{-1} = \int_p D_{\mu\nu}~ \tau\gamma^\mu ~G ~\tau\gamma^{\nu}~.
\end{eqnarray}

The first step that we take is to commute the first $\tau$ and $\gamma^\mu$ in (\ref{SDap}), so that in the middle of the integrand 
we have a matrix product given by
\begin{eqnarray}\label{tgt}
\tau G \tau &=& X\begin{pmatrix} e_1 & -i\epsilon \\ i\epsilon & e_2 \end{pmatrix} \begin{pmatrix} \slashed{p}-m_2 & \mu \\
\mu & \slashed{p}-m_1 \end{pmatrix}\begin{pmatrix} e_1 & -i\epsilon \\ i\epsilon & e_2 \end{pmatrix}\\
&=& X \begin{pmatrix}
         e_1^2(\slashed{p}-m_2)+\epsilon^2(\slashed{p}-m_1) & -Y+\mu(e_1 e_2 - \epsilon^2) \\
        Y+\mu(e_1 e_2 - \epsilon^2) & \epsilon^2(\slashed{p}-m_2)+e_2^2(\slashed{p}-m_1)
         \end{pmatrix}~, \nonumber
\end{eqnarray}
where
\begin{eqnarray}
X &=& i\frac{p^2 + \slashed{p}( m_1 + m_2) + m_1 m_2 -\mu^2}{(p^2 - m_1^2)(p^2 - m_2^2)-2 \mu^2(p^2 + m_1 m_2)+\mu^4}\\
  &=& i\frac{p^2 + \slashed{p}( m_1 + m_2) + m_1 m_2 -\mu^2}{(p^2 - A_{-}^2)(p^2-A_{+}^2)}~;\nonumber\\
Y &=& i \epsilon[ e_1 (\slashed{p}-m_2)+ e_2 (\slashed{p}-m_1)]~,\nonumber
\end{eqnarray}
with $A_{\pm}^2$ defined as in (\ref{Apmdef}). 
If we identify individually each matrix element in the Schwinger-Dyson equation (\ref{SDap}), we obtain for the $M_{11}$ element
\begin{eqnarray}
i m_1 &=& \int_p D_{\mu\nu}~\gamma^\mu X [e_1^2(\slashed{p}-m_2)+\epsilon^2(\slashed{p}-m_1)] \gamma^{\nu}\\
     &=&\int_p \frac{(4+\zeta)}{(1+ \vec{p}^2/M^2) } \frac{p^2(e_1^2 m_1 + \epsilon^2 m_2) 
     + (\mu^2-m_1 m_2)(e_1^2 m_2 +\epsilon^2 m_1)}{p^2(p^2 - A_{-}^2)(p^2-A_{+}^2)}~,\nonumber
\end{eqnarray} 
The last equation can be written as
\begin{eqnarray}\label{m1}
\frac{m_1}{4+\zeta} =(e_1^2 m_1 + \epsilon^2 m_2) I_1 + (\mu^2-m_1 m_2)(e_1^2 m_2 +\epsilon^2 m_1)I_2,
\end{eqnarray}
where
\begin{eqnarray}\label{I1I2mink}
I_1 &=& -i \int_p \frac{1}{1+\vec{p}^2/M^2} \frac{1}{(p^2 - A_{-}^2)(p^2-A_{+}^2)}\\
I_2 &=& -i \int_p \frac{1}{(1+\vec{p}^2/M^2) } \frac{1}{p^2(p^2 - A_{-}^2)(p^2-A_{+}^2)}~.\nonumber
\end{eqnarray} 
The Wick rotation $p_0 \rightarrow i\omega$ leads to 
\begin{eqnarray}\label{i1i2in}
I_1 &=& \frac{1}{4\pi^3} \int_0^{\infty} \frac{\vec{p}^2d\vec p}{1 + \vec{p}^2/M^2} \int_{-\infty}^{\infty}
\frac{d\omega}{(\omega^2+ \vec{p}^2 + A_{+}^{2})(\omega^2+ \vec{p}^2 + A_{-}^{2})}\\
I_2&=&\frac{-1}{4\pi^3} \int_0^{\infty}\frac{\vec{p}^2d\vec p}{1 + \vec{p}^2/M^2} \int_{-\infty}^{\infty} 
\frac{d\omega}{(\omega^2+\vec{p}^2)(\omega^2+ \vec{p}^2 + A_{+}^{2})(\omega^2+ \vec{p}^2 + A_{-}^{2})}~.\nonumber
\end{eqnarray}
The integrand of $I_1$ which depends on $\omega$ only can be written
\bea\label{I1t}
&&\frac{1}{A_{+}^2-A_{-}^2}\left[\left(\frac{1}{\omega^2+\vec{p}^2}-\frac{1}{(\omega^2+ \vec{p}^2 + A_{+}^{2})}\right)\right.\nn
&&~~~\left.~~~~~~~~~~~~~~-\left(\frac{1}{\omega^2+\vec{p}^2}-\frac{1}{(\omega^2+ \vec{p}^2 + A_{-}^{2})}\right) \right]~,
\eea
and similarly, the integrand of $I_2$ which depends on $\omega$ only can be expressed as
\bea\label{I2t}
&&\frac{1}{A_{+}^2-A_{-}^2}\left[\frac{1}{A_{-}^2}\left(\frac{1}{\omega^2+\vec{p}^2}-\frac{1}{(\omega^2+ \vec{p}^2 + A_{-}^{2})}\right)\right.\nn
&&\left.~~~~~~~~~~~~~~~~~~~~~~~~-\frac{1}{A_{+}^2}\left(\frac{1}{\omega^2+\vec{p}^2}-\frac{1}{(\omega^2+ \vec{p}^2 + A_{+}^{2})}\right) \right]
\eea
Finally, substituting (\ref{I1t}) and (\ref{I2t}) into (\ref{i1i2in}) leads to the first equation of (\ref{4equs}).\\
Furthermore, due to the symmetry of our model, the second equation of (\ref{4equs}) is obtained from the first one by exchanging $m_1$ and $m_2$.\\
Finally, the left-hand side of (\ref{SDap}) is symmetric with non-diagonal elements given by $i \mu$, therefore, the non-diagonal 
elements of the right-hand side must also be equal. However, looking at (\ref{tgt}), we realize that it is only possible if the terms 
related with $Y$ vanish. So, the non-diagonal elements give us the following equations
\begin{eqnarray}
i \mu &=&\int_p \frac{(4+\zeta)}{(1+ \vec{p}^2/M^2) } \mu(e_1 e_2 -\epsilon^2) \frac{p^2  + m_1 m_2 -\mu^2}{p^2(p^2 - A_{-}^2)(p^2-A_{+}^2)}~,\\
0 &=& \int_p D_{\mu\nu}~\gamma^\mu X Y \gamma^{\nu}\nonumber\\
&=& \epsilon \int_p \frac{(4+\zeta)}{(1+ \vec{p}^2/M^2) } \frac{p^2(e_1 m_1 + e_2 m_2) 
+ (\mu^2-m_1 m_2)(e_1 m_2 +e_2 m_1)}{p^2(p^2 - A_{-}^2)(p^2-A_{+}^2)}~,\nonumber
\end{eqnarray}
where using eq.(\ref{I1I2mink}), leads to the last two equations (\ref{4equs}).

\section{Appendix: one-loop fermion self energy}

We calculate here, in the Feynman gauge, the fermion wave function renormalization for the case \{$e_1=e_2$ and $\epsilon=0$\}. 
In order to avoid IR divergences obtained in the one-loop calculation for $m_1=m_2=\mu$, because one of the eigen masses vanishes,
we consider here the situation $m_1=m_2=m\ne\mu$. The fermion propagator is then given by
\begin{eqnarray}\label{fprop}
G(p)=i\frac{p^2+2m\slashed{p}+m^2-\mu^2}{[p^2-(m+\mu)^2][p^2-(m-\mu)^2]}\begin{pmatrix} \slashed{p}-m  & \mu \\ \mu & \slashed{p}-m \end{pmatrix}.
\end{eqnarray}
We obtain the fermion wave function renormalization by differentiating the fermion self-energy with respect to the external momentum and then, set it to zero. 
Since the fermion propagator (\ref{fprop}) has two independent flavour components, we consider the one-loop diagonal self energy $\Sigma^{(1)}_{diag}$ 
and the one-loop off-diagonal part $\Sigma^{(1)}_{off}$, where
\begin{eqnarray}
&&\Sigma^{(1)}_{diag} (\omega,\vec{p})\\ 
&=& \frac{-i e^2}{(2\pi)^4} \int\frac{d^4 k}{1+\vec{k}^2/M^2} \left\{\frac{\gamma^\mu \gamma_\mu [(p-k)^2 - (m^2-\mu^2)]m}
{k^2 [(p-k)^2-(m+\mu)^2][(p-k)^2-(m-\mu)^2]}\right.\nn
&&~~~~~~~~~~~+\left.\frac{\gamma^\mu(\slashed{p}-\slashed{k})\gamma_\mu[(p-k)^2-(m^2+\mu^2)] }{k^2 [(p-k)^2-(m+\mu)^2][(p-k)^2-(m-\mu)^2]}\right\}\nn
&&\Sigma^{(1)}_{off}(\omega, \vec{p}) \nn
&=& \frac{-i e^2}{(2\pi)^4} \int\frac{d^4 k}{1+\vec{k}^2/M^2} \frac{ \gamma^\mu\gamma_\mu [(p-k)^2+m^2-\mu^2]\mu 
+2 m \mu\gamma^\mu(\slashed{p}-\slashed{k})\gamma_\mu }{ k^2 [(p-k)^2-(m+\mu)^2][(p-k)^2-(m-\mu)^2]}~.\nonumber
\end{eqnarray}
Differentiating these terms with respect to $p_\rho$ and then setting $\omega = 0$ and $\vec{p} =0$, we find
\begin{eqnarray}
\frac{\partial \Sigma^{(1)}_{diag}}{\partial p_\rho}\big|_{p=0} 
&=& \frac{i e^2}{8 \pi^4} \int\frac{d^4 k}{1+\vec{k}^2/M^2} \left\{ \frac{k^2 \gamma^\rho - (m^2 +\mu^2) \gamma^\rho 
+ 2k^\rho \slashed{k}}{k^2[k^2-(m+\mu)^2][k^2-(m-\mu)^2]} \right.\\
&& - \left.\frac{ 4 k^\rho \slashed{k} k^4 -8 k^\rho \slashed{k} k^2 (m^2+\mu^2) +4k^\rho \slashed{k} 
(m^2+\mu^2)^2}{k^2[k^2-(m+\mu)^2]^2[k^2-(m-\mu)^2]^2}\right\}\nn
\frac{\partial \Sigma^{(1)}_{off}}{\partial p_\rho} \big|_{p=0} 
&=& - \frac{i \mu m e^2}{4 \pi^4}\int\frac{d^4 k}{1+\vec{k}^2/M^2} \left\{ \frac{-\gamma^\rho}{ k^2 [k^2-(m+\mu)^2][k^2-(m-\mu)^2]} \right.\nn
&&  + \left.\frac{ 4 k^\rho \slashed{k}k^2 - 4 k^\rho \slashed{k}(m^2+\mu^2) }{k^2[k^2-(m+\mu)^2]^2[k^2-(m-\mu)^2]^2}\right\}~.\nn
\end{eqnarray}
We write then
\bea
\Sigma^{(1)}_{diag}&=&-m+Z_{diag}^0\omega\gamma^0-Z_{diag}^1\vec p\cdot\vec\gamma\nn
\Sigma^{(1)}_{off}&=&-\mu+Z_{off}^0\omega\gamma^0-Z_{off}^1\vec p\cdot\vec\gamma~,
\eea
and since we are interested in the limit $\mu\to m$, we write $m-\mu=m\delta$, with $\delta<<1$ and approximate $m + \mu \approx 2 m$.
In terms of new variables $x=\sqrt{ \vec{k}^2 }/m$, $y = k_0/m$, $\lambda = m/M\ll 1$ and after a Wick rotation, we obtain
\begin{eqnarray}
Z_{diag}^0 &=& \frac{e^2}{2 \pi^3}\int_{0}^{\infty} \frac{x^2 dx}{1+\lambda^2 x^2} \int_{-\infty}^{\infty} dy
\left[ \frac{-(x^2+y^2)-2y^2-2}{(x^2+y^2)(x^2+y^2+4)(x^2+y^2+\delta^2)}\right. \nn
&& +\left. \frac{4 y^2 (x^2+y^2)^2 +16y^2(x^2+y^2)+16y^2}{(x^2+y^2)(x^2+y^2+4)^2(x^2+y^2+\delta^2)^2}\right]\\
Z_{off}^0 &=& \frac{e^2}{\pi^3}\int_{0}^{\infty} \frac{x^2 dx}{1+\lambda^2 x^2}  \int_{-\infty}^{\infty} dy
\left[ \frac{1}{(x^2+y^2)(x^2+y^2+4)(x^2+y^2+\delta^2)}\right.\nn
&& -\left. \frac{4 y^2 (x^2+y^2) + 8 y^2}{(x^2+y^2)(x^2+y^2+4)^2(x^2+y^2+\delta^2)^2}\right]~,\nonumber
\end{eqnarray}
and 
\begin{eqnarray}
Z^1_{diag} &=& \frac{e^2}{2 \pi^3}\int_{0}^{\infty} \frac{x^2 dx}{1+\lambda^2 x^2} \int_{-\infty}^{\infty} dy
\left[ \frac{-(x^2+y^2)-2x^2/3-2}{(x^2+y^2)(x^2+y^2+4)(x^2+y^2+\delta^2)}\right. \nn
&& +\left. \frac{4}{3}\frac{ x^2 (x^2+y^2)^2 +4x^2(x^2+y^2)+4x^2}{(x^2+y^2)(x^2+y^2+4)^2(x^2+y^2+\delta^2)^2}\right]\\
Z^1_{off} &=& \frac{e^2}{\pi^3}\int_{0}^{\infty} \frac{x^2 dx}{1+\lambda^2 x^2}  \int_{-\infty}^{\infty} dy
\left[ \frac{1}{(x^2+y^2)(x^2+y^2+4)(x^2+y^2+\delta^2)}\right.\nn
&& -\left. \frac{4}{3}\frac{x^2 (x^2+y^2) + 2 x^2}{(x^2+y^2)(x^2+y^2+4)^2(x^2+y^2+\delta^2)^2}\right]~.\nonumber
\end{eqnarray}
Finally, after solving these integrals, we find 
\begin{eqnarray}
Z^0_{diag} &=& \frac{e^2}{8 \pi^2}\left(\frac{1}{4}-\frac{1}{2} \ln{2} + \frac{1}{2} \ln{\delta} + \ln{\lambda}   \right) \\
Z^0_{off} &=& \frac{e^2}{16 \pi^2}\left(\ln{2} - \ln{\delta} \right)~,  \nonumber
\end{eqnarray}
and
\begin{eqnarray}
Z^1_{diag} &=&\frac{e^2}{8 \pi^2}\left(-\frac{1}{12}-\frac{1}{2} \ln{2} + \frac{1}{2} \ln{\delta} + \ln{\lambda}   \right)\\
Z^1_{off} &=& \frac{e^2}{16 \pi^2}\left(\ln{2} - \ln{\delta} \right)~.  \nonumber
\end{eqnarray}

\end{document}